\documentclass{iet-ell}
\makeatother
\usepackage{amsmath,amsfonts}
\usepackage{algorithmic}
\usepackage{algorithm}
\usepackage{array}
\usepackage[caption=false,font=normalsize,labelfont=sf,textfont=sf]{subfig}
\usepackage{textcomp}
\usepackage{stfloats}
\usepackage{url}
\usepackage{verbatim}
\usepackage{bm}
\usepackage{xcolor}
\usepackage{todonotes}
\usepackage{booktabs}
\usepackage{tabularx}
\usepackage{xspace}
\usepackage{tikz}
\newcommand{\reuse}{$\texttt{MLA}_\texttt{ru}$\xspace}
\newcommand{\recompute}{$\texttt{MLA}_\texttt{rc}$\xspace}
\newcommand{\derived}{$\texttt{MHA}_\texttt{l}$\xspace}
\newcommand{\scaled}{$\texttt{MHA}_\texttt{s}$\xspace}

\newcommand{\circled}[1]{\tikz[baseline=(char.base)]{
    \node[shape=circle,draw,inner sep=0.5pt] (char) {\tiny #1};}}

\begin{document}

\title{Hardware-Centric Analysis of DeepSeek’s Multi-Head Latent Attention}

\author[af1]{Robin~Geens~}
\orcid{0009-0000-2450-2660}
\author[af1]{Marian~Verhelst}
\orcid{0000-0003-3495-9263}
\affil[af1]{MICAS, KU Leuven, Leuven, Belgium}
\corresp{Email: robin.geens@kuleuven.be}

\begin{abstract}
Multi-Head Latent Attention (MLA), introduced in DeepSeek-V2, improves the efficiency of large language models by projecting query, key, and value tensors into a compact latent space. This architectural change reduces the KV-cache size and significantly lowers memory bandwidth demands, particularly in the autoregressive decode phase.
This letter presents the first hardware-centric analysis of MLA, comparing it to conventional Multi-Head Attention (MHA) and evaluating its implications for accelerator performance. We identify two alternative execution schemes of MLA—reusing, resp. recomputing latent projection matrices—which offer distinct trade-offs between compute and memory access. Using the Stream design space exploration framework, we model their throughput and energy cost across a range of hardware platforms and find that MLA can shift attention workloads toward the compute-bound regime.

Our results show that MLA not only reduces bandwidth usage but also enables adaptable execution strategies aligned with hardware constraints. 
Compared to MHA, it provides more stable and efficient performance, particularly on bandwidth-limited hardware platforms. 
These findings emphasize MLA’s relevance as a co-design opportunity for future AI accelerators.
\end{abstract}

\maketitle

\section{\textbf{Introduction}}
DeepSeek-V3~\cite{deepseekv3} has been shown to significantly reduce training and inference costs compared to other commercial large language models, while maintaining competitive accuracy and usability. A key enabler of this efficiency is its use of Multi-Head Latent Attention (MLA), a novel attention mechanism where $Q$, $K$ and $V$ matrices are first projected into a low-dimensional latent space, and then projected into a higher-dimensional space to compute the attention scores. This approach allows for compact storage of the KV-cache entries during inference, drastically reducing memory bandwidth requirements, particularly in the decode stage.

This letter presents a hardware-centric analysis of MLA’s decode-phase behavior on modern accelerators, comparing its performance to that of traditional Multi-Head Attention (MHA). The analysis quantifies the associated throughput and energy cost improvements, and evaluates the resulting shift in architectural requirements for efficient deployment.
Although previous works have detailed the benefits of MLA as an algorithmic technique \citep{review-of-techniques,trans-mla,ji2025economicalinferenceenablingdeepseeks}, to the best of our knowledge, this is the first study of its kind to analyze the computational footprint and practical implications on hardware acceleration systems.

\begin{figure}[tb]
    \centering
    \includegraphics[width=\linewidth]{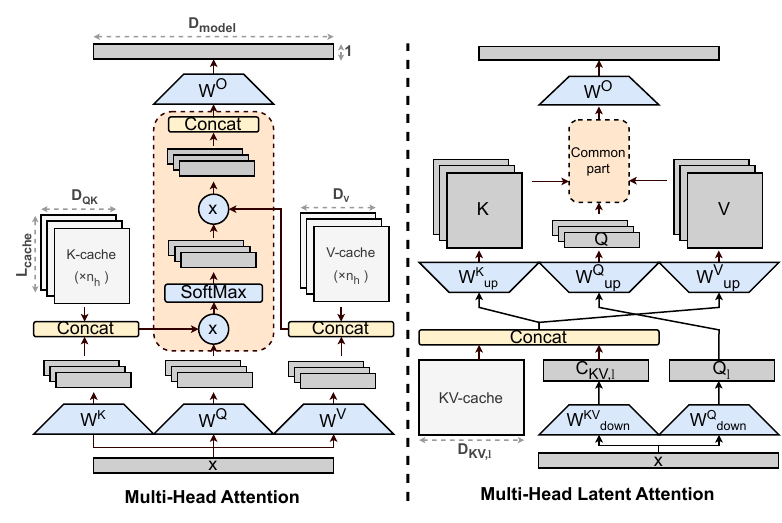}
    \caption{Architecture of MHA and MLA.
    }
    \label{fig:model-arch}
\end{figure}

\section{\textbf{Organization}}
This letter begins with a review of standard MHA and the key modifications introduced in MLA.
We then analyze the ordering of matrix multiplications in MLA, identifying trade-offs between compute and memory access. Building on these insights, we compare operation counts, memory access patterns, and algorithmic intensities of MHA and MLA. Finally, we model these characteristics across various hardware platforms using the Stream design space exploration framework to derive implications for accelerator architecture design.

\subsection{\textbf{Multi-Head Attention}}
MHA~\citep{vaswani2023attentionneed}, shown in Figure~\ref{fig:model-arch} (left) is defined as follows:
\begin{gather*}
    \text{MHA}(X) = \text{Concat}(\text{head}_1, \dots, \text{head}_{n_h}) \; W^O   
\end{gather*}
where each attention head is computed as:
\begin{gather*}
    \text{head}_i = \text{SoftMax} \left( \frac{Q_iK_i^T}{\sqrt{D_{QK}}} \right) V_i = S_i \; V_i
\end{gather*}
\vspace{-3mm}

\begin{gather*}
    Q = X \; W^Q; \quad K=X \; W^K; \quad V=X \; W^V \\ 
    X \in \mathbb{R}^{L \times D_{\text{model}}} \\
    W^Q, W^K \in \mathbb{R}^{n_h \times D_{\text{model}} \times D_{QK}}; \quad W^V \in \mathbb{R}^{n_h \times D_{\text{model}} \times D_V}\quad \\
    W^O \in \mathbb{R}^{n_h D_V \times D_{\text{model}}} \\
\end{gather*}
\vspace{-4mm}

\noindent $Q_i$, $K_i$ and $V_i$ are obtained from slicing the $Q$, $K$ and $V$ matrices into $n_h$ equal parts in the $D$-dimension. During autoregressive inference, $K$ and $V$ matrices are cached and updated incrementally as each new token is generated.

\subsection{\textbf{Multi-Head Latent Attention}} 
In MLA~\citep{deepseekv2} (Figure~\ref{fig:model-arch} (right)), inputs $X$ are first projected into a smaller, latent space\footnote{For simplicity, Rotary Positional Embeddings (RoPE)~\cite{rope} is omitted here.}:
\begin{gather*}
    Q_l = X \; W_{\text{down}}^Q; \quad C_{KV,l}=X \; W_{\text{down}}^{KV} \\
    Q = Q_l \; W_{\text{up}}^Q; \quad K=C_{KV,l} \; W_{\text{up}}^K; \quad V=C_{KV,l} \; W_{\text{up}}^V \\
\end{gather*}
\vspace{-8mm}

where
\begin{gather*}
    W_{\text{down}}^Q \in \mathbb{R}^{D_{\text{model}} \times D_{Q,l}} \quad W_{\text{down}}^{KV} \in \mathbb{R}^{D_{\text{model}} \times D_{KV,l}} \\
    W_{up}^Q \in \mathbb{R}^{n_h \times D_{Q,l} \times D_{QK}}; \quad W_{\text{up}}^K \in \mathbb{R}^{n_h \times D_{KV,l} \times  D_{QK}}; \\
    \quad W_{\text{up}}^V \in \mathbb{R}^{n_h \times D_{KV,l} \times D_V}
\end{gather*}

The computational benefits of this approach are twofold, assuming $D_{Q_l}, D_{KV_l} \ll D_{QK}$: 1)~the $C_{KV_l}$ matrix is cached instead of the large $K$ and $V$ matrices, significantly reducing the memory footprint and 2)~the number of parameters in the projection weights is much smaller. 
 
\begin{table}[h]
    \centering
    \caption{Parameters of DeepSeek-V3~\citep{deepseekv3} model and derived variants}
    \label{tab:params}
    \begin{tabular}{c|ccc}
        \hline
        \textbf{Parameter} & \textbf{MLA} & \textbf{MHA (derived)}  & \textbf{MHA (scaled)}\\
        \hline
        $D_{\text{model}}$ & 7168 & 7168 & 4363 \\
        $n_h$              & 128  & 128  & 128 \\
        $D_{Q,l}$          & 1536 & -    & -   \\
        $D_{KV,l}$         & 512  & -    & -   \\
        $D_{QK}$           & 128  & 128  & 77  \\
        $D_{V}$            & 128  & 128  & 77  \\
        \begin{tabular}[c]{@{}c@{}}\#params in 1\\attention layer\end{tabular} & 174M & 470M & 172M \\
        \hline
    \end{tabular}
\end{table}

In this letter, we analyze MLA with the hyperparameter instantiations proposed in DeepSeek-V3 and given in Table~\ref{tab:params}. To compare MLA with standard MHA, we propose two MHA baselines: one with equivalent internal dimensions yet larger number of parameters (\derived) and another with an equivalent parameter count (\scaled).

\subsection{\textbf{Order of Multiplications}}
Computing the attention scores requires a projection from the cache's latent space to the $K$- and $V$-spaces. A naive implementation of MLA would up-project the entire cached latent history before computing attention. However, this can be avoided by reordering operations:
\begin{align*}
    Z = Q K^T &= (Q_l \; W_{\text{up}}^Q)(C_{KV,l} \;  W_{\text{up}}^K)^T \\
   &= Q_l 
    \underset{\circled{1}}{} W_{\text{up}}^Q 
    \underset{\circled{2}}{} W_{\text{up}}^{K,T} 
    \underset{\circled{3}}{} C_{KV,l}^T
\end{align*}

\begin{figure}
    \centering
    \includegraphics[width=0.8\linewidth]{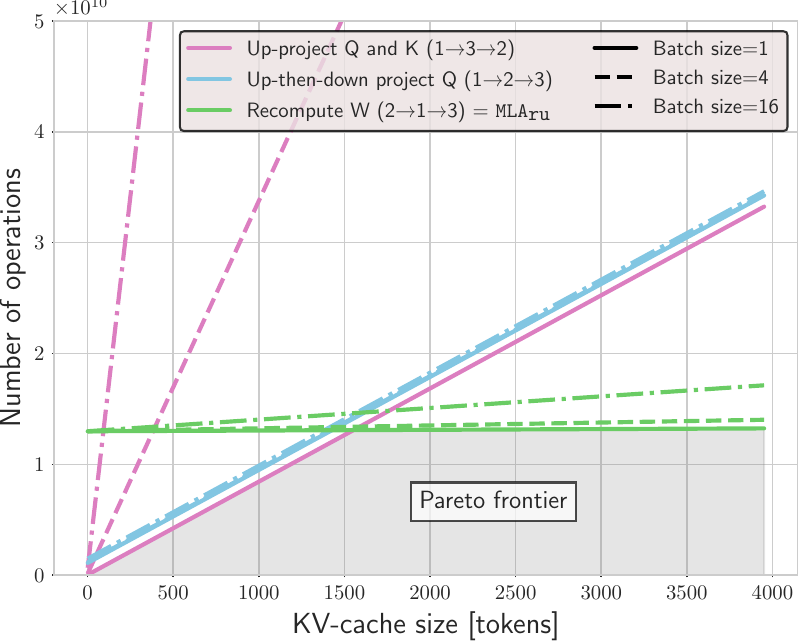}
    \caption{Required number of operations for different computation orders of $Q_l \; W_{\text{up}}^Q 
 \; W_{\text{up}}^{K,T} \; C_{KV,l}^T$ in the DeepSeek-V3 decode phase. \texttt{1→2→3} indicates left-to-right multiplication. For typical and high sequence length scenarios, first recomputing the absorbed weight matrix and transforming $Q_l$ to the $KV,l$-space results in the least amount of operations.}
    \label{fig:QK-order}
\end{figure}

\noindent The order in which these matrix multiplications are performed has a significant impact on efficiency.
A naive strategy that computes the leftmost and rightmost products first (\texttt{1→3→2}) is suboptimal, as it requires up-projecting the entire latent KV-cache before performing attention in the high-dimensional embedding dimension.
A left-to-right ordering (\texttt{1→2→3}) incrementally transforms $Q_l$ first to the full space and then to the $K_l$-space, deferring the attention computation to the $K_l$-space and reducing compute and bandwidth costs. 
Another alternative is to compute the middle product first (\texttt{2→1→3}), which directly transforms $Q_l$ into the $K_l$-space. The approach of first computing the composite matrix $W_{\text{up}}^Q W_{\text{up}}^{K,T}$ is known as \emph{weight absorption}~\citep{sinai2025mla}, and this computation order is referred to as \recompute in the remainder of this letter. As shown in Figure~\ref{fig:QK-order}, the \recompute ordering generally yields the best performance, particularly for long KV-caches and small batch sizes.

A similar analysis can be made for the output projection:
\begin{gather*}
    Y = S \; V\; W^O = S \; C_{KV,l} \; W_{\text{up}}^V  \; W^O
\end{gather*}
In this case, executing right-to-left is typically most efficient.

\subsection{\textbf{Recompute vs. Reuse Trade-Off}} 
Instead of recomputing $W_{up}^Q W_{up}^{K,T}$ at each inference step, the absorbed weight matrix can also be precomputed and reused. This modifies the previous approach to $QK^T= Q_l \; W_{\text{absorb}} \; K^T$ with left-to-right execution order. We refer to this variant as \reuse.
Given that $D_{Q,l} D_{KV,l} > D_{Q,l} D_{QK} + D_{QK} D_{KV,l}$, \reuse saves computation but requires more memory bandwidth compared to \recompute. MLA thus offers a built-in mechanism to trade computations for memory accesses depending on hardware constraints.

Continuing on previous insights, the remainder of this letter will analyze four alternative attention methods: 1) \reuse with precomputation of the absorbed weight matrix; 2) \recompute with on-the-fly weight recomputation; 3) \derived : a regular MHA variant with identical $D_{\text{model}}$ but more parameters; and 4) \scaled : an MHA variant scaled-down to match MLA's parameter count (Table~\ref{tab:params}).

\subsection{\textbf{Operations and Memory Accesses}}

\begin{figure}[tb]
    \centering
    \includegraphics[width=\linewidth]{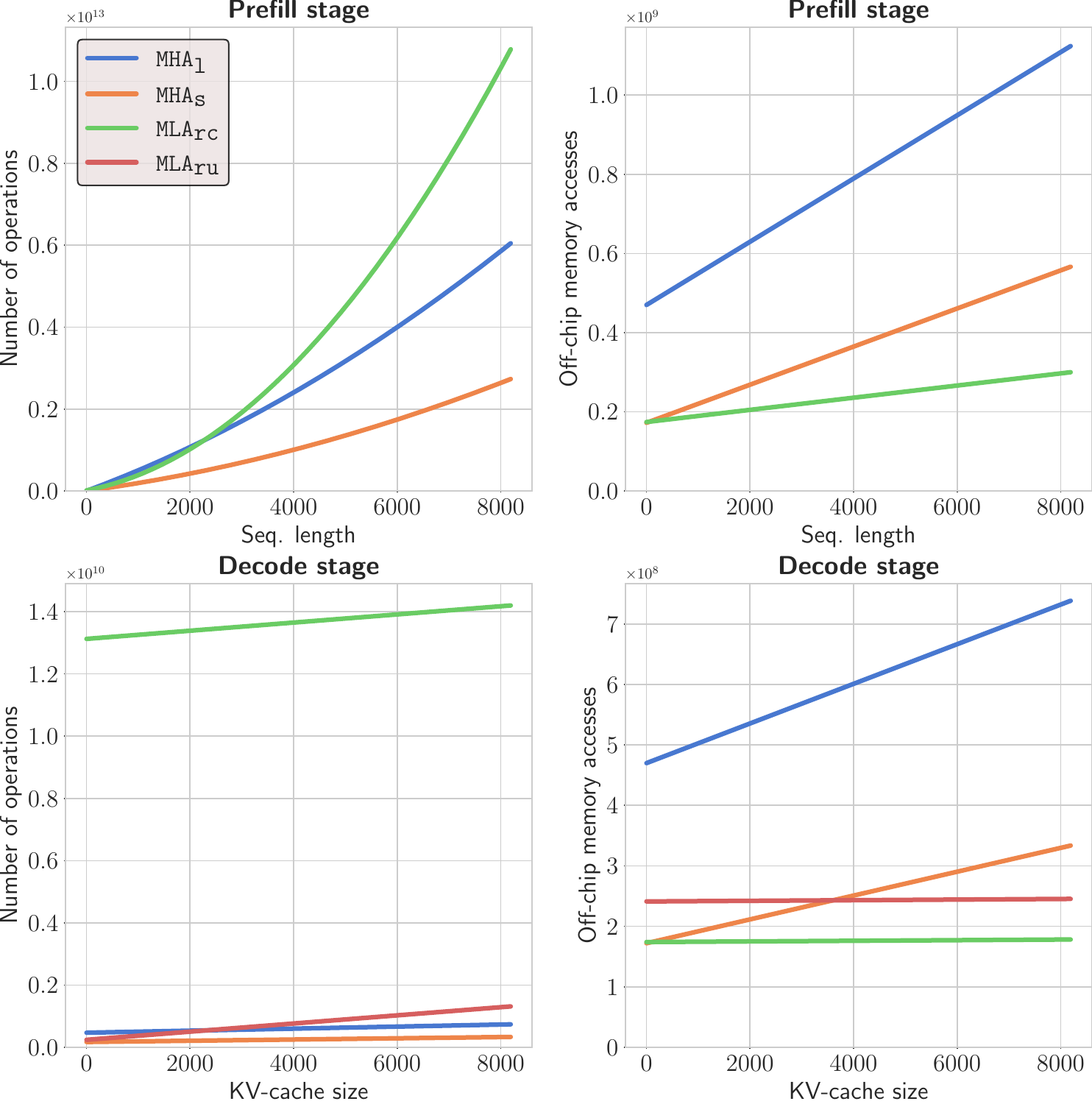}
    \caption{Number of operations and number of external memory accesses for a single attention layer (batch size = 1). MLA uses the \textit{Recompute W (\texttt{2→1→3})} multiplication order.
    }
    \label{fig:compute-stats}
\end{figure}

Figure~\ref{fig:compute-stats} compares the total number of operations and off-chip memory accesses for the four attention methods during prefill and decode stages.
We assume that all computations can be performed without additional memory accesses of intermediate activations - an assumption that will be validated in the next section. 
The number of accesses for \scaled and \recompute starts out equal, but \recompute scales better for larger sequences due to the smaller cache dimension. Overall, \recompute trades additional computations for reduced memory accesses.

\begin{figure}[tb]
    \centering
    \includegraphics[width=\linewidth]{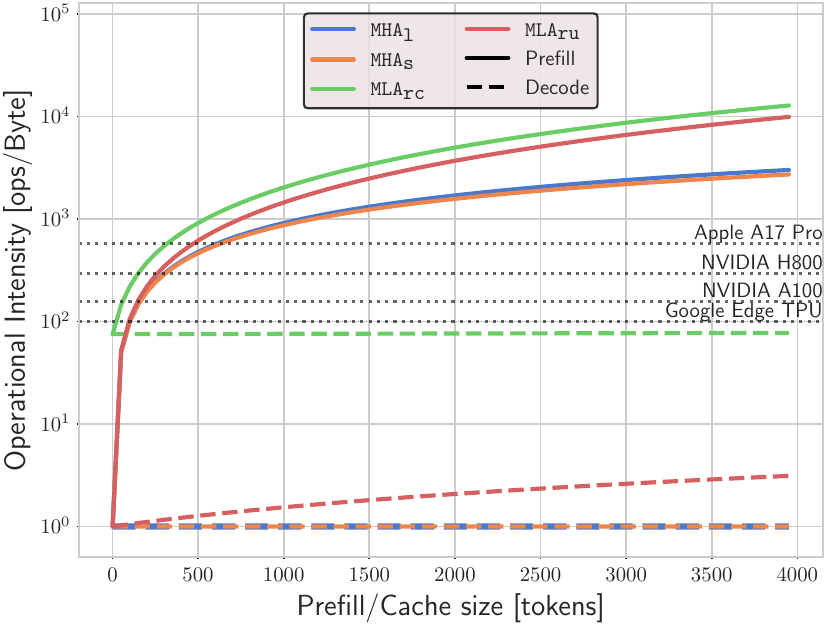}
    \caption{Operational Intensity (OI) of attention methods in function of sequence length (prefill phase) or KV-cache size (decode phase). Dotted lines indicate the roofline corner points (i.e., the OI that marks the transition from memory-bound to compute-bound) of well-known platforms.}
    \vspace{-6mm}
    \label{fig:intensities}
\end{figure}

To analyze and compare performance, it is essential to examine the operational intensities (OI), defined as the total number of operations divided by the number of off-chip memory accesses. This metric helps determine whether the workload is compute-bound or memory-bound.
Figure~\ref{fig:intensities} shows the OIs (in operations/byte) of the four methods, at varying sequence lengths (prefill stage), resp. KV-cache length (decode stage). All methods exhibit a high OI in the prefill stage, due to the large number of required computations and possibility to reuse weights across multiple token vectors. In the decode stage, however, there is a notable difference between the assessed attention methods. Both \derived and \scaled maintain a consistently low OI regardless of KV-cache size. In contrast, the OI of \reuse strongly depends on the KV-cache size, as the number of operations scales linearly with the cache size while the marginal cost of latent cache entries is insignificant compared to the constant size of the weight matrices. Meanwhile, \recompute exhibits a significantly higher OI with a minimal sensitivity to cache size. This is because the constant computational cost of recomputing the weight matrix dominates over the relatively minor cost of cache-size dependent projections. 

Although all four methods remain memory-bound during the decode phase on the commercial platforms shown in Figure~\ref{fig:intensities}, they exhibit substantially different OI. Consequently, each method’s relative performance depends on the platform’s roofline corner. For example, the \recompute method’s much higher OI allows it to nearly reach the roofline corner of a compute-limited device like the Google Edge TPU. In contrast, this same OI falls well below the roofline corner of the more compute-rich Apple A17 Pro. Because of these differing OI characteristics, no single method universally outperforms the others across all platforms. This variation motivates analyzing performance across a range of hardware configurations to account for different compute/memory trade-offs.
The remainder of this letter focuses on the single-batched decode stage, where this trade-off plays the most prominent role. Moreover, this stage is typically the bottleneck in contemporary hardware platforms and especially in real-time applications.

\section{\textbf{Hardware Modeling with Stream}}
To quantify the relative benefits of each attention method, we model their execution on hardware platforms with varying characteristics using Stream~\citep{stream}, a design space exploration (DSE) framework tailored for estimating and optimizing the performance of multi-core dataflow accelerators. Stream ingests an accelerator architecture description and a target workload as inputs, based on which it models on-chip dataflow, memory hierarchy, and inter-core connections under hardware constraints. This allows the tool to analytically estimate bandwidth usage, energy consumption and inference latency for a given workload on the specified hardware architecture.

To ensure broadly applicable insights, we adopt a generalized AI accelerator architecture as a reference, which consists of a spatial 2D array of MAC units, a vector unit with non-linear function units, a unified on-chip memory, and a design-time configurable off-chip bandwidth, modeled after~\citep{kao2022flatoptimizeddataflowmitigating}. When recomputing $W_{\text{up}}^Q W_{\text{up}}$, it is crucial that the resulting, larger weight matrix remains on-chip. Otherwise, the benefit of recomputation is entirely lost. For this purpose, we configure Stream to execute the matrix multiplications in a fused manner. Note that Stream also models the Softmax execution, which was neglected in Figure~\ref{fig:compute-stats}.

\begin{figure}[tb]
    \centering
    \includegraphics[width=\linewidth]{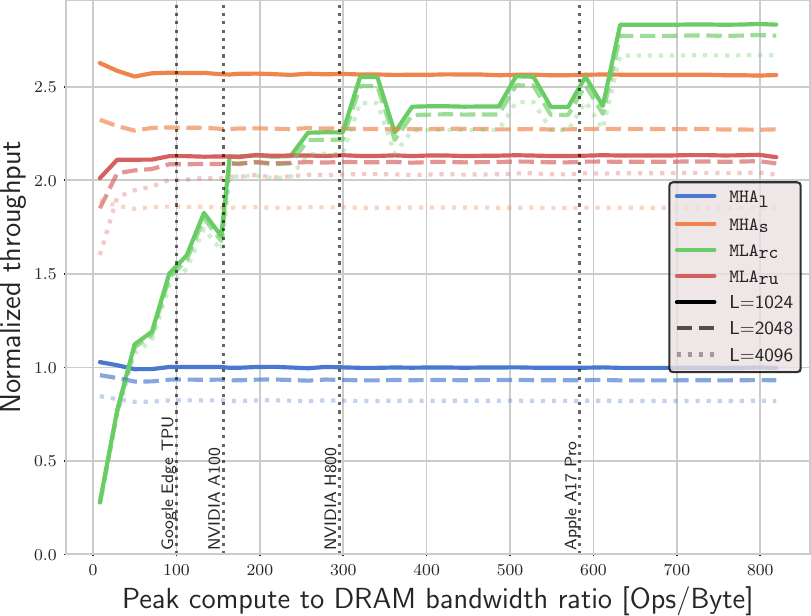}
    \caption{Stream estimated throughput of a single attention layer in function of peak compute over DRAM bandwidth ratio at a constant bandwidth of 400 GB/s (batch=1). Dotted lines indicate the roofline corner points of well-known platforms.
    }
    \vspace{-6mm}
    \label{fig:latency_stream}
\end{figure}

\begin{figure}[tb]
    \centering
    \includegraphics[width=\linewidth]{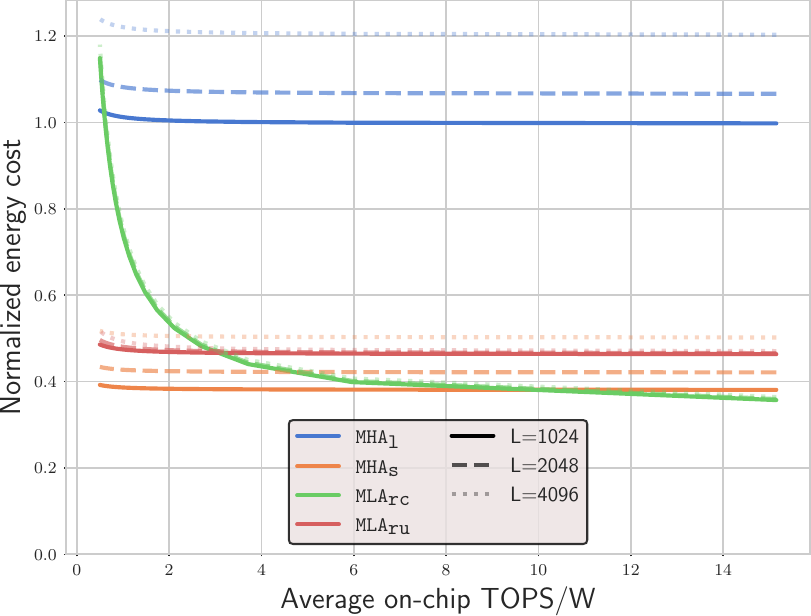}
    \caption{Stream estimated energy for a single attention layer in function of the average on-chip TOPS/W at constant $E_{\text{DRAM,bit}}$ = 8 pJ (batch=1).
    }
    \vspace{-6mm}
    \label{fig:energy_stream}
\end{figure}

\section{\textbf{Performance Analysis}}
Since our primary interest lies in comparing the benefits and overheads of the four attention methods for a range of hardware configurations, we explore their relative performance as a function of the hardware platform's compute-to-bandwidth ratio, expressed in terms of peak operations per second over peak off-chip memory bandwidth. Figure~\ref{fig:latency_stream} summarizes the resulting layer throughput performance in function of this compute-to-bandwidth ratio, evaluated across three KV-cache sizes.

Among the methods, \recompute results in the highest relative performance, benefiting from reduced memory transfers at the cost of additional arithmetic, except for the cases where the accelerator has little compute resources available compared to its off-chip bandwidth. In this uncommon case, it is more beneficial to reuse the weight matrix and reload it from DRAM at each iteration. Note that both \reuse and \recompute implement the same algorithm with identical weights; the choice between them can be made dynamically based on deployment constraints and hardware capabilities.

Although \scaled can approach the performance of \recompute for small cache sizes, this advantage quickly diminishes with larger caches. In general, the performance of MHA is highly sensitive to the length of the previously computed KV-cache, due to high dimensionality of cache entries. In contrast, MLA exhibits much more stable performance across varying cache sizes, making it easier to ensure consistent quality-of-service under different sequence lengths and runtime conditions.

\section{\textbf{Energy Analysis}}
Based on average costs per operation, Stream also provides an estimated energy cost per inference.
To assess relative energy efficiency across attention methods under divergent OI characteristics,
we again focus on two key hardware parameters: the accelerator's on-chip efficiency, expressed in $E_{\text{op}}$ or TOPS/W, and $E_{\text{DRAM,bit}}$. The latter depends on the used DRAM technology and is typically a design constraint.
Figure~\ref{fig:energy_stream} presents the resulting normalized energy estimates for varying accelerator efficiencies. 
While the performance analysis identified \recompute as the best-performing method for typical hardware characteristics, this conclusion does not universally extend to energy usage and instead depends heavily on the platform's characteristics. In contrast, \reuse is much more resistant to changes in the hardware characteristics. Additionally, although \scaled can be the most energy efficient for some hardware design points, this only holds for small KV cache sizes and the spread on MHA's results is once again significantly larger.

\section{\textbf{Conclusion}}
This letter presented a hardware-oriented analysis of Multi-Head Latent Attention (MLA) in DeepSeek-V3, focusing on its decode-phase behavior. By projecting activations into a low-dimensional latent space, MLA significantly reduces off-chip memory traffic, leading to higher operational intensity and making it better suited to compute-bound accelerators.

Using the Stream design space exploration framework, we evaluated MLA against two baselines based on the standard Multi-Head Attention (MHA) formulation, and explored two MLA variants—\recompute and \reuse—that offer trade-offs between compute and memory usage. \recompute consistently achieved the highest throughput and intensity across a range of hardware models, while \reuse proved advantageous on platforms with limited compute resources. In contrast, MHA variants remained memory-bound and showed greater performance sensitivity to cache size and hardware configuration. Overall, our results show that MLA enables adaptable attention execution tailored to hardware characteristics. This flexibility makes it particularly promising for future AI accelerators, where balancing compute and bandwidth remains a critical design challenge.

\begin{acks}
This project has been partly funded by the European Research Council (ERC) under grant agreement No. 101088865, the European Union’s Horizon 2020 program under grant agreement No. 101070374, the Flanders AI Research Program, Research Foundation Flanders (FWO) under grant No. 1S37125N, and KU Leuven.
\end{acks}

\balance

\bibliographystyle{iet}
\bibliography{ref.bib}	

\end{document}